\documentclass[sigplan,10pt]{acmart}

\setcopyright{none}
\renewcommand\footnotetextcopyrightpermission[1]{}

\usepackage{multirow}
\usepackage{tcolorbox}
\usepackage{comment}
\tcbuselibrary{breakable,skins}
\usepackage{placeins} 
\usepackage{enumitem}
\usepackage{subcaption}
\usepackage{natbib}
\usepackage[utf8]{inputenc} 
\usepackage[T1]{fontenc}    
\usepackage{hyperref}       
\usepackage{cleveref}       
\usepackage{url}            
\usepackage{booktabs}       
\usepackage{amsfonts}       
\usepackage{nicefrac}       
\usepackage{microtype}      
\usepackage{xspace}
\usepackage{amsmath}
\usepackage[dvipsnames]{xcolor}

\usepackage{listings} 

\newcommand{\eat}[1]{} 

\newif\ifshowtasks
\showtasksfalse   

\ifshowtasks
  \newcommand{\task}[1]{\textcolor{blue}{\textbf{[Task: #1]}}}
\else
  \newcommand{\task}[1]{}
\fi

\newcommand{\after}[1]{[{\color{red}AFTER: #1}]}
\renewcommand{\after}[1]{} 

\usepackage{minted}

\lstdefinestyle{mystyle}{
    backgroundcolor=\color{backcolour},   
    commentstyle=\color{codegreen},
    keywordstyle=\color{magenta},
    numberstyle=\tiny\color{codegray},
    stringstyle=\color{codepurple},
    basicstyle=\ttfamily\footnotesize,
    breakatwhitespace=false,         
    breaklines=true,                 
    captionpos=b,                    
    keepspaces=true,                 
    numbers=left,                    
    numbersep=5pt,                  
    showspaces=false,                
    showstringspaces=false,
    showtabs=false,                  
    tabsize=2
}

\newcommand{\sys}{StateFork\xspace}
\newcommand{\substrate}{Waypoint\xspace}

\begin{document}

\date{}

\title{\sys: Branchable Infrastructure\\ for Agent Exploration}


\author{Jiakai Xu}
\authornote{Both authors contributed equally to this research.}
\affiliation{%
  \institution{Columbia University}
  \country{USA}
}
\email{ax2155@columbia.edu}
\orcid{0009-0006-2074-9109}

\author{Tianle Zhou}
\authornotemark[1]
\affiliation{%
  \institution{Columbia University}
  \country{USA}
}
\email{mz2998@columbia.edu}

\author{Georgios Liargkovas}
\affiliation{%
  \institution{Columbia University}
  \country{USA}
}
\email{g.liargkovas@columbia.edu}
\orcid{0000-0002-8896-4044}

\author{Danielle Gillai}
\affiliation{%
  \institution{Columbia University}
  \country{USA}
}
\email{deg2184@barnard.edu}

\author{Ruizhe Fu}
\affiliation{%
  \institution{Google}
  \country{USA}
}
\email{rf2715@columbia.edu}

\author{Patrick Shen}
\affiliation{%
  \institution{Columbia University}
  \country{USA}
}
\email{pts2125@columbia.edu}

\author{Eugene Wu}
\affiliation{%
  \institution{Columbia University}
  \country{USA}
}
\email{ew2493@columbia.edu}
\orcid{0000-0003-4254-6688}

\author{Kostis Kaffes}
\affiliation{%
  \institution{Columbia University}
  \country{USA}
}
\email{kkaffes@cs.columbia.edu}
\orcid{0000-0002-0517-7206}



\begin{abstract}
AI agents improve task success by exploring multiple trajectories, but for computer-use agents each trajectory modifies external environment state.
Branching from an intermediate point is correct only when restoration is observation-equivalent---future actions produce the same observations---and practical only when creating, restoring, and discarding branch states is physically efficient.
We study this problem for terminal-using agents, where tasks modify files, shell context, running processes, and local services.
We introduce \sys, a logical control plane that separates exploration policies from physical state materialization, exposing sessions, commands, snapshots, restores, and cleanup over multiple execution substrates.
We also build \substrate, a checkpoint/restore substrate for terminal execution sessions that combines filesystem layering, process checkpointing, and a persistent terminal-compatible command session.
Together, \sys and \substrate improve terminal-agent exploration by combining sample-efficient search with efficient restoration of the right execution state.
On Terminal-Bench, branch-based exploration through \sys and \substrate improves task completion over pass@$20$ at the same visited-node budget, and using \substrate achieves 26\% higher task accuracy than other execution substrates while completing exploration up to 70\% faster.
These results show that observation-equivalent, physically efficient execution sessions are a key systems abstraction for exploratory AI agents.
\end{abstract}

\settopmatter{printfolios=true}
\maketitle
\pagestyle{plain}

\section{Introduction}

AI agents increasingly solve tasks by exploring.
They test hypotheses, retry failed actions, compare alternative plans, and backtrack from unproductive continuations before committing to a final answer.
For reasoning-only tasks, exploration is mostly a logical operation.
An agent can branch by copying a prompt, a reasoning trace, or a small scratchpad.
For computer-use agents, however, exploration is also a systems problem.
Each action changes an external environment, and future observations depend on the state created by earlier actions.
Branching therefore requires the agent runtime to manage not only the model context, but also the environment state associated with intermediate points in the trajectory.

The correctness requirement for such branching is \emph{observation equivalence}.
After restoring an intermediate node, subsequent actions should produce observations equivalent to those produced from the original node.
This constraint is weaker than bit-for-bit equality of the host machine as the execution substrate need not preserve state that no future command uses.
It is also stronger than restoring only a convenient, albeit incomplete, subset of the necessary state.
If a branch omits state that affects future observations, or if effects from one speculative branch contaminate another, the search tree no longer represents valid alternatives.
Successful exploration also requires \emph{physical efficiency}.
Creating, restoring, and discarding branches must be cheap enough that exploration improves task performance rather than being dominated by state-management overhead.

In this paper, we focus on enabling such exploration for terminal-use agents.
Terminal interaction is a natural target because many practical coding and computer-use agents operate through command-line interfaces, including OpenAI Codex CLI~\citep{openai_codex_cli} and Claude Code~\citep{anthropic_claude_code}.
The terminal is text-native, compositional, and already exposes many tools needed for software engineering tasks, enabling agents to inspect directories, edit files, install dependencies, run tests, start services, and debug failures through command output.
At the same time, these actions modify state that is not captured by the command transcript alone.
Throughout the paper, we use Terminal-Bench~\citep{tbench_2025} as a representative workload for terminal-use agents because it uses this same command--observation interface and consists of a wide variety of generic computer-use tasks.

We call the relevant state an \emph{execution session}: the agent-facing computer environment produced by the command prefix leading to a node in the exploration tree.
For terminal agents, an execution session includes the state that future shell commands can observe or depend on.
This includes filesystem state such as source files, installed packages, virtual environments, generated artifacts, logs, and configuration; terminal-session state such as the working directory, exported variables, activated environments, aliases, shell functions, job-control state, and PTY-dependent behavior; process and memory state such as running programs, REPLs, servers, background jobs, open file descriptors, and in-memory caches; and local service state such as sockets, databases, and loopback services.
Our manual labeling of Terminal-Bench tasks shows that these categories are common rather than exceptional.
All labeled tasks require filesystem state, most require terminal-session state, and a nontrivial fraction require live process and memory state.

Existing execution substrates do not provide the right combination of observation equivalence and physical efficiency.
Replay can reconstruct a node by re-executing the prefix from the root, but its cost grows with prefix length and it is fragile under nondeterminism.
Undo can be efficient when every action has a reliable inverse, but terminal actions such as installing packages, mutating databases, deleting files, or starting services are not generally reversible.
Filesystem-only checkpoints lose shell and process state; process checkpoints assume a compatible filesystem and session boundary; container checkpoints pay deployment-oriented lifecycle and image-management costs; VM snapshots preserve too much machine state; and branch-oriented APIs provide fast speculation for single applications but not restorable checkpoints over execution sessions.
Further, adopting any of these mechanisms requires the agent to understand the implementation details and low-level operations in order to save and restore state.
This limits portability to new execution substrates that are faster, more resource efficient, and/or better preserve observation equivalence for a class of desired tasks.

\begin{figure}[t]
  \centering
  \includegraphics[width=0.95\linewidth]{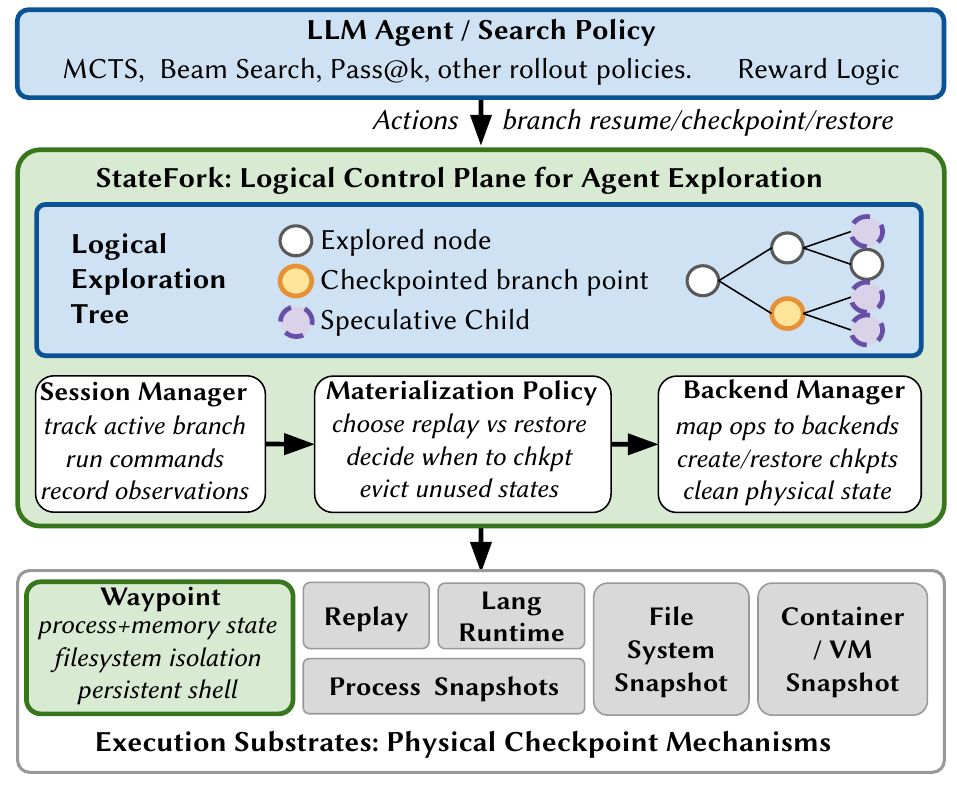}
  \caption{\textbf{Terminal-agent stack (our work in green).}
An agent or search policy expands a logical tree over command--observation trajectories.
\sys provides the logical control plane: it exposes branchable sessions, chooses whether branch points are physical or virtual, and maps logical operations to substrate-specific restore mechanisms.
Execution substrates materialize logical branch points using mechanisms such as replay-to-node, filesystem snapshots, and container/VM snapshots; \substrate is our optimized substrate for terminal agents, capturing the task-facing execution-session state: filesystem, process/memory, and persistent terminal-session state.}
  \label{fig:statefork-overview}
  \vspace{-2pt}
\end{figure}

We introduce two layers to address this gap.
Figure~\ref{fig:statefork-overview} shows the resulting stack.
At the top, an agent or search policy expands a logical exploration tree over command--observation trajectories.
In the middle, \sys acts as the logical control plane.
It exposes branchable session operations, tracks logical branch points, decides how those branch points should be materialized, and dispatches backend-specific operations.
At the bottom, execution substrates implement the physical restoration mechanism, including replay-to-node, filesystem snapshots, container or VM snapshots, and our own substrate, \substrate.

\sys separates the logical exploration tree from physical state materialization, so the same exploration algorithm can run over different execution substrates while the runtime optimizes the tradeoff between observation equivalence and physical efficiency.
\substrate is the checkpoint/restore substrate we built for terminal-agent execution sessions.
It treats the execution session as the native checkpoint boundary.
Filesystem layering preserves file contents, process checkpointing preserves application memory, and a persistent terminal-compatible command session preserves shell context.
\substrate is full-state with respect to the execution-session boundary that determines future terminal observations, not with respect to the entire host machine or a portable deployment artifact.

Our evaluation shows that \sys and \substrate improve both sides of terminal-agent exploration.
First, branch-based exploration through \sys improves sample efficiency.
On Terminal-Bench, Monte Carlo Tree Search achieves a higher task completion rate than pass@$20$ for the same visited-node budget, showing that reusing intermediate nodes is more effective than repeatedly restarting from the root.
Second, \sys backed by \substrate improves both sample and physical efficiency by efficiently restoring observation-equivalent execution sessions.
In microbenchmarks, \substrate achieves up to $10\times$ lower checkpoint-plus-restore latency than existing VM and container substrates that capture both memory and filesystem state.
In macrobenchmarks, \sys with \substrate achieves 26\% higher task completion rate than other evaluated platforms while completing exploration 70\% faster than replaying commands from the initial state.
Together, \sys and \substrate let agents express exploration as a logical tree while restoring the terminal execution sessions needed for observation-equivalent and physically efficient branching.

In summary, this paper makes the following contributions:
\begin{itemize}[leftmargin=1em]
  \item We formalize terminal-agent exploration as the \emph{observation-equivalent and efficient restoration} of intermediate nodes.
  \item We define the \emph{execution session} as the task-facing state boundary needed for terminal-agent observation equivalence, and show that Terminal-Bench tasks commonly require filesystem, terminal-session, process, memory, and local-service state.
  \item We introduce \emph{\sys}, a logical control plane that lets exploration algorithms execute commands, create branch points, restore prior nodes, and optimize how those logical nodes are materialized across execution substrates.
  \item We design and evaluate \emph{\substrate}, an execution substrate that captures execution-session state efficiently using file system layering, process checkpointing, and a persistent terminal-compatible command session.
  \item We show that \sys and \substrate together improve the Terminal-Bench task completion rate by >26\% with >70\% lower latency.
\end{itemize}

\section{Fundamentals of Agent Exploration}
\label{sec:motivating_example}

\subsection{Logical Agent Exploration}
\label{sec:logical-agent-exploration}

Language-model agents increasingly solve tasks by exploring multiple possible trajectories before committing to one.
We use \emph{logical agent exploration} to mean the search problem faced by the agent that decomposes to two subproblems: deciding (i) which unfinished trajectory prefixes to expand, and (ii) which continuation to pursue.
This logical problem is independent of the mechanism used to materialize the corresponding environment states.

Formally, let $I$ be the task instructions,
$\tau_{<t}=(a_0,o_0,\ldots, \allowbreak a_{t-1},o_{t-1})$ be the trajectory prefix observed by the agent where $a_i$ and $o_i$ are the action taken and outcome observed at the $i$-th step, respectively, and let $M_t$ denote any framework-managed memory included in the model context.
At step $t$, the agent chooses an action
\[
  a_t =
  \mathrm{LLM}(I,\tau_{<t},M_t).
\]
In a terminal environment, $a_t$ is a command or tool invocation.
Executing $a_t$ against the current environment state $S_t$ produces an observation and a successor state:
\[
  (o_t,S_{t+1}) = \llbracket a_t \rrbracket(S_t).
\]

The central objective of logical agent exploration is \emph{sample efficiency}.
Each LLM call chooses an action, expands a trajectory prefix, and consumes part of the exploration budget.
Because long-horizon tasks have many possible continuations, an effective exploration method should focus on calls  most likely to change the final outcome rather than repeatedly sample the same uninformative prefixes.

\begin{figure}[t]
  \centering
  \includegraphics[width=0.95\linewidth]{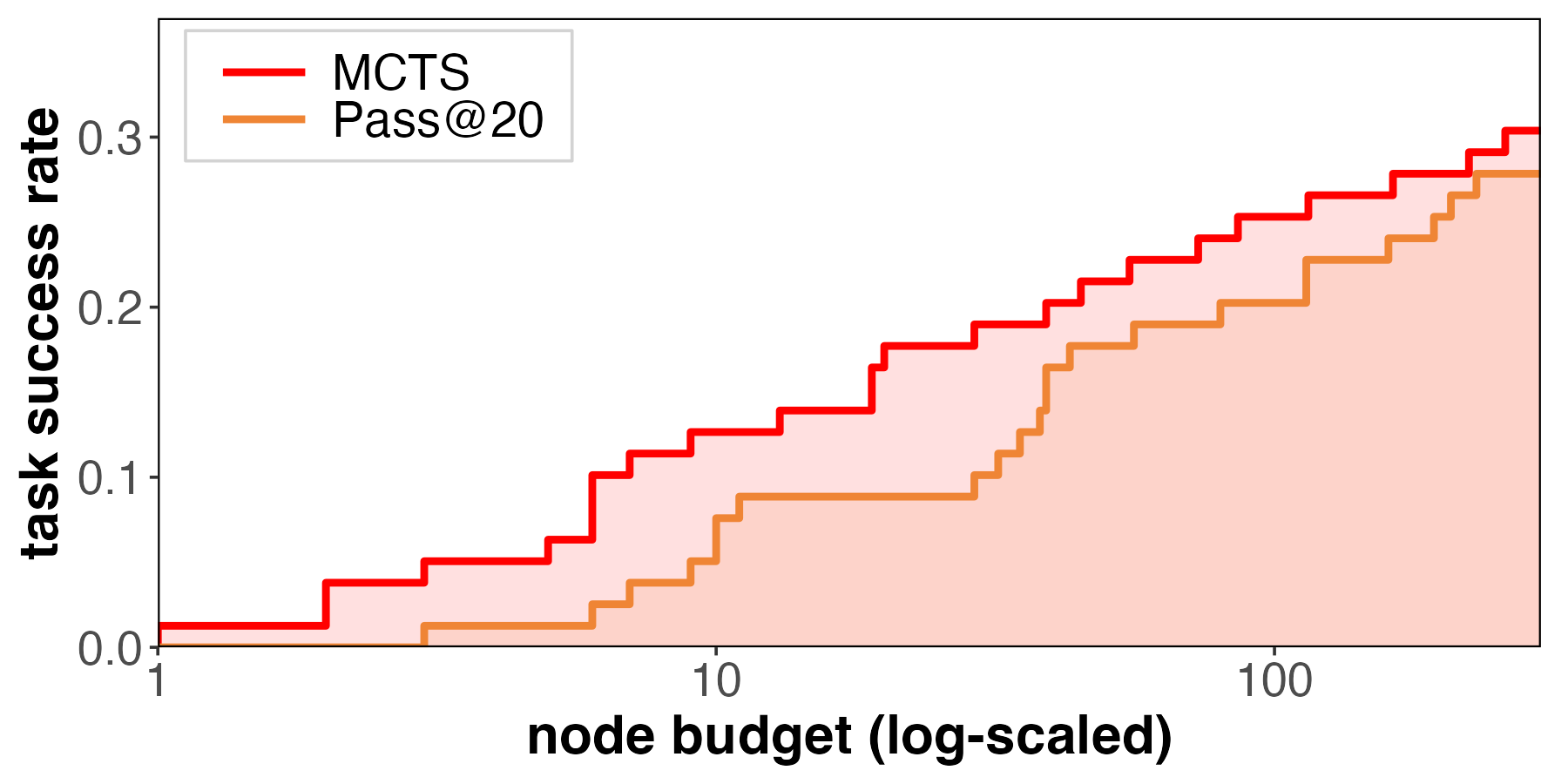}
  \caption{Terminal-Bench task success as a function of visited search nodes, where each node corresponds to an LLM call.
  MCTS achieves higher success than pass@$20$ at the same node budget because it reuses observations and intermediate environment states across sibling branches.}
  \label{fig:tbench-nodes-success}
\end{figure}

The most naive method is pass@$k$: run the agent from the initial state up to $k$ independent times, and count the task as solved if any rollout succeeds.
Pass@$k$ is simple and often improves over a single attempt, but it is not sample efficient.
Each rollout restarts from the root, so intermediate trajectory prefixes are not reused as branch points.
Intermediate nodes improve sample efficiency in two ways.
First, the prefix leading to a useful node is sampled once and then reused for multiple continuations.
Second, observations at that node provide evidence about which nearby branches are promising, allowing later calls to be allocated adaptively.
This is why recent exploration methods represent agent search as a tree: the root is the initial state, edges are actions, and nodes are partial trajectories that can be expanded selectively.
Tree of Thoughts~\citep{yao2023treethoughtsdeliberateproblem}, LATS~\citep{zhou2024languageagenttreesearch}, Tree Search for Language Model Agents~\citep{koh2026treesearchlanguagemodel}, and ExACT\citep{yu2025exactteachingaiagents} all use this branch-and-reuse structure to spend LLM calls on promising partial trajectories rather than independent restarts.

Figure~\ref{fig:tbench-nodes-success} provides additional evidence for this distinction for Terminal-Bench~\citep{tbench_2025}, the most prominent benchmark of terminal-based tasks in which agents solve problems by issuing shell commands and observing their outputs.
At the same visited-node budget, Monte Carlo Tree Search (MCTS) achieves higher Terminal-Bench task completion rate than pass@$20$.
Each additional call can refine an existing search tree instead of beginning another isolated rollout, improving sample efficiency.

\subsection{Physical Requirements for Agent Exploration}
\label{sec:physical-requirements}

Logical exploration assumes that the agent can branch from an intermediate node in the search tree.
For pure reasoning tasks, this assumption is almost free.
The relevant state is usually the prompt, the reasoning trace, and any structured scratchpad maintained by the search algorithm.
Forking a node means copying tokens or a small in-memory object.
This is one reason tree-based exploration has been effective in reasoning settings.

Terminal-use agents are different.
A node in the search tree is not only a transcript prefix.
It is also the computer state produced by executing that prefix.
To make branch-based exploration correct and practical, the execution substrate must satisfy two physical requirements: \emph{observation equivalence} and \emph{physical efficiency}.

\paragraph{Observation equivalence.}
The correctness condition for revisiting a node is observation equivalence.
Fix the agent context
$x_t=(I,\tau_{<t},M_t)$.
Two states $S$ and $S'$ are observation-equivalent for this context if every future action sequence the agent may take from $x_t$ produces equivalent observations from $S$ and $S'$:
\[
  S \equiv_{x_t} S'
  \quad\Longleftrightarrow\quad
  \forall \alpha \in \mathcal A(x_t),\;
  \mathrm{Obs}(S,\alpha) \simeq \mathrm{Obs}(S',\alpha).
\]
Observation equivalence is weaker than bit-for-bit equality of the host machine.
The runtime does not need to preserve state that no future command can observe.
However, it is stronger than restoring only a convenient subset of state.
If omitted state can affect future observations, then the restored node is not the same node for purposes of exploration.
The same requirement also implies containment since effects from one speculative branch must not leak into another unless they belong to a common ancestor.
For example, packages installed, package-manager metadata mutated, files generated, or services started in one branch must not contaminate observations in a sibling branch.

For terminal-use agents, the relevant state is an \emph{execution session}: the agent-facing environment created by the command prefix leading to a node.
This includes the state that future shell commands can observe or depend on.
Filesystem state includes source files, installed packages, virtual environments, configuration, generated artifacts, logs, and task data.
Terminal session state includes the working directory, exported variables, activated environments, aliases, shell functions, job-control state, and PTY-dependent behavior.
Process and memory state includes running programs, REPLs, servers, background jobs, open file descriptors, and in-memory application state.
Local service state includes sockets, databases, loopback services, and other service-local state inside the task environment.
External state, such as clocks, package registries, remote APIs, and the public internet, is not generally restorable by a local runtime; preserving equivalence for such effects requires caching, proxying, mocking, or treating them as nondeterministic inputs.

Table~\ref{tab:state-requirements} shows that these categories are not incidental in Terminal-Bench.
All labeled tasks require filesystem state, most require terminal-session state, and a nontrivial fraction require live process, memory, or local-service state.
Thus, a terminal-agent branch point cannot generally be represented by the reasoning trace, nor even a filesystem snapshot alone.

\begin{table}[t]
\centering
\small
\setlength{\tabcolsep}{4pt}
\caption{Task-relevant state in manually labeled Terminal-Bench tasks. Categories are not mutually exclusive.}
\label{tab:state-requirements}
\begin{tabular}{@{}p{0.28\linewidth}p{0.42\linewidth}cc@{}}
\toprule
\textbf{State boundary} & \textbf{Examples} & \textbf{Count} & \textbf{Pct.} \\
\midrule
File contents / filesystem
& Application files, installed packages, generated data, configuration
& 77 & 100.00\% \\
Terminal session
& Working directory, environment variables, activated virtual environments, shell-local context
& 52 & 67.53\% \\
Application memory / process state
& Running programs, servers, REPLs, background jobs, in-memory state
& 13 & 16.88\% \\
Local service / network state
& Sockets, local services, network-dependent task state
& 11 & 14.29\% \\
\bottomrule
\end{tabular}
\end{table}

\paragraph{Physical efficiency.}
Observation equivalence is necessary but not sufficient.
Exploration also requires \emph{physical efficiency} as creating, restoring, and discarding branch states must be cheap relative to the LLM-call budget they are meant to support.
A perfectly faithful restoration mechanism is not useful for fine-grained search if every branch requires rebuilding the environment from scratch, copying a large filesystem, restarting all services, or constructing a full deployment artifact.
Physical efficiency has several dimensions.
Restore latency affects how quickly the search algorithm can expand another branch.
Storage overhead determines how many speculative states can be kept available.
The substrate should therefore preserve exactly the task-facing execution-session state needed for observation equivalence, while avoiding unnecessary work for state that cannot affect the agent's future observations.

These two requirements are coupled.
If the preserved state is too narrow, restoration is fast but incorrect.
If the preserved state is too broad, restoration may be correct but too expensive for search.
The goal is therefore not whole-machine equality, but efficient restoration of the minimal practical execution-session state that determines future observations for terminal-use agents.

\subsection{Existing Execution Substrates Are Insufficient}
\label{sec:substrate-mismatch}

The previous subsection identifies two requirements for terminal-agent exploration:
restoring intermediate nodes must preserve observation equivalence, and doing so must be physically efficient.
There are three broad ways to implement such restoration.

\paragraph{Replay-to-node}
resets to the initial state and re-runs the action prefix to the target node.  While this provides observation equivalence under deterministic execution and reproducible external effects, its cost grows with the length and cost of the action prefix, and penalizes deeper trajectories.    Sources of non-determinism (e.g., time, randomness, network) also affect equivalence.  While simple, it is not a reliable substrate.  

%

\paragraph{Undo} relies on having inverse or compensating actions~\citep{chang2025sagallmcontextmanagementvalidation}.  
Terminal actions, however, often do not have faithful inverses: file deletion, package installation, database mutation, permission updates, and service creation can all incur side effects that are difficult to invert.  Undo actions may be partial, and fail to restore hidden process state, service-local state, or shell context.


\paragraph{Checkpoint/restore}
is a natural primitive for branch-based exploration.
It directly records and restores the state of an intermediate node; it avoids replay and does not require invertible actions.  
The challenge is choosing the right checkpoint scope and capturing it efficiently: too narrow and restoration is fast but violates observation equivalence, too broad and restoration is observation-equivalent but slow.  
For terminal-use agents, the scope is an execution session: filesystem state, terminal-session state, process and memory state, and relevant local services.

Existing execution substrates provide pieces of this abstraction, but none provides both observation equivalence and physical efficiency at the execution-session boundary.

\paragraph{Filesystem checkpoints.}
Filesystem checkpointing captures durable file state.
Layered filesystems such as ZFS~\citep{bonwick2003zfs}, Btrfs~\citep{rodeh2013btrfs}, OverlayFS~\citep{linux_overlayfs}, UnionFS~\citep{quigley2006unionfs}, and BranchFS~\citep{wang2026forkexplorecommitos} provide efficient copy-on-write views and can preserve source edits, installed packages, generated artifacts, and configuration files.
Transactional filesystems and storage systems such as Warp~\citep{escriva2016warp}, QuickSilver~\citep{schmuck1991quicksilver}, Stasis~\citep{sears2006stasis}, and transactional file access via kernel extensions~\citep{spillane2009transactional} support atomic file commit or rollback.
These file-centered mechanisms can lose shell-local context, activated environments, running servers, REPLs, background jobs, PTYs, and in-memory state, so they do not by themselves restore the complete execution session.

\paragraph{Process checkpoints and forks.}
Process mechanisms capture volatile execution state.
Zap/CRAK~\citep{osman2002zap,zhong2001crak}, DMTCP~\citep{ansel2009dmtcp}, and
Faasm~\citep{shillaker2020faasm} are earlier process-checkpointing systems across kernel, userspace, and distributed-function settings respectively.
More recently introduced, CRIU~\citep{criu} can dump and restore a process tree, including memory, registers, file descriptors, and process metadata.
Process-forking mechanisms such as \texttt{fork()}~\citep{linux_fork} are even cheaper, but they duplicate only the calling process at the time of the fork.
Similarly, \texttt{branch()}-style APIs~\citep{wang2026forkexplorecommitos} are useful when all relevant work is scoped to a particular process or branch context, but they do not automatically capture an arbitrary terminal session that has accumulated a shell, descendants, PTYs, mounts, files, and local services over time.
Raw process checkpointing is therefore complementary to filesystem checkpointing as it preserves important live state, but it assumes that the restored process tree sees a compatible filesystem, mount layout, runtime filesystems, PTYs, devices, sockets, and command interface.
It is necessary for observation equivalence, but not sufficient as a standalone execution-session checkpoint.

\paragraph{Checkpointable runtimes and notebooks.}
A separate line of work makes higher-level runtimes checkpointable.
Notebook systems such as Kishu~\citep{li2024kishu}, ElasticNotebook~\citep{li2023elasticnotebook}, Multiverse Notebook~\citep{sato2024multiverse}, and Fork It~\citep{weinman2021forkit} support time travel, live migration, or stateful alternatives by tracking notebook variables, cell dependencies, interpreter state, or notebook branches.
Runtime-integrated systems such as OpenJDK CRaC~\citep{openjdk_crac} and CheckSync~\citep{kaashoek2022checksync} coordinate checkpointing with the Java and Go runtimes, while continuation-based systems such as Kappa~\citep{zhang2020kappa} and Stackless Python~\citep{stackless_python_pickling} serialize language-level execution state.
These systems show the value of raising checkpointing above raw OS snapshots, but their boundaries are typically a language heap, notebook kernel, function continuation, or initialized runtime and are therefore too narrow for general terminal-agent exploration.

\paragraph{VM and container snapshots, commits, and sandboxes.}
Containers and VMs can capture broader boundaries.
Container commits preserve filesystem changes, while Docker \citep{docker} and Podman~\citep{podman,podman-checkpoint,podman-restore} checkpointing can also capture CRIU process images.
MicroVM snapshots, such as Firecracker snapshots~\citep{firecracker,firecracker-snap}, preserve guest memory and device-model state; gVisor~\citep{gvisor,moby-issue,gvisor-issue} strengthens the application-kernel boundary under sandbox-specific constraints.
MBOX \citep{kim2013mbox} and try~\citep{lamprou2026try, liargkovas2023executing} mediate or selectively apply command effects, while Sysfilter~\citep{demarinis2020sysfilter}, SysXCHG~\citep{gaidis2023sysxchg}, and Draco~\citep{skarlatos2020draco} prevent unsafe system calls.
These mechanisms target deployment, migration, isolation, effect control, or filtering, not arbitrary restore of terminal-session nodes; VM/container variants are also too coarse and costly~(\S~\ref{sec:microbenchmark}).
Thus, they do not expose the terminal execution session---PTY state, shell context, local services, and runtime mounts---as the native branch/restore boundary.

\paragraph{This paper.}
These gaps motivate a separation between logical exploration and physical materialization.
\S~\ref{sec:statefork} presents \sys, the logical control plane that lets agents create branch points, restore prior nodes, and clean up speculative branches without committing to one restoration mechanism.
Because logical nodes are explicit, \sys can run and optimize the same exploration policy over different backends.
\S~\ref{sec:waypoint} presents \substrate, our execution-session substrate for terminal agents.
\substrate combines filesystem layering, process checkpointing, and a persistent terminal-compatible command session to restore the state needed for observation-equivalent terminal-agent exploration with high physical efficiency.


\section{\sys: Logical Control Plane for Agent Exploration}
\label{sec:architecture}
\label{sec:statefork}

\sys separates the logical search tree used by an agent from the physical mechanisms used to restore environment state.
Logical exploration decides which trajectory prefixes are worth expanding.
Physical materialization decides how the environment state for those prefixes is represented, restored, and eventually discarded.
The purpose of \sys is to connect these layers while keeping their concerns separate: the agent or search policy operates over logical branch points, while execution substrates implement the physical restoration of those branch points.

\paragraph{How an agent uses \sys.}
An agent interacts with \sys through a small set of concrete session operations:
A search policy calls \texttt{build\_session(backend, config)} to start a live execution session.
It then calls \texttt{exec\_command(cmd)} to issue an action and receive an observation.
At any point, \texttt{snapshot()$\to$id} names the current trajectory prefix as a logical branch point, \texttt{restore(id)}returns to that prior branch point, and \texttt{cleanup(id)} discards a branch or subtree that is no longer needed.
For example, an MCTS policy can restore a parent node, execute a candidate command sequence, score the resulting observation, snapshot the child node, restore the parent again, and expand a sibling branch.
The search policy sees these as operations on a logical tree, while the runtime may implement each \texttt{snapshot()} or \texttt{restore(id)} via replay, filesystem snapshots, a container or VM checkpoint, or \substrate.

\sys in \Cref{fig:statefork-overview} has three components: a session manager, a materialization policy, and backend managers.

\subsection{Session Manager}
\label{sec:statefork-api}

The session manager owns one live execution session and a tree of named logical branch points.
It exposes agent-facing operations to build an environment, run commands, name branch points, restore prior nodes, and clean up discarded branches.
It also records the metadata common to all substrates: parent pointers, children, command logs, timing measurements, storage usage, the current active node, and whether each logical node is physical or virtual.

This interface gives exploration algorithms a stable target.
MCTS, beam search, or other exploration policies transparently use different restoration algorithms and execution substrates, including replay-to-node, filesystem snapshots, container checkpoints, VM snapshots, hybrid process/filesystem checkpoints, and our own substrate, \substrate.
\sys can leverage substrate-specific features---e.g., checkpoint scope, compression, export mode, or physical-versus-virtual materialization--on behalf of the agent search policy.
\sys also lets us compare substrates and optimizations for the same control flows in an apples-to-apples manner. 

The key abstraction is the separation between \emph{logical branch points} and \emph{backend materialization}.
A logical branch point is the node that the agent or search algorithm wants to revisit.
A physical checkpoint is only one possible implementation of that node.
This distinction matters because exploration algorithms often create many more logical nodes than they eventually revisit.
Without this separation, the runtime must either materialize every node eagerly or force the search algorithm to reason directly about backend-specific checkpoint costs.

\subsection{Materialization Policy}
\label{sec:statefork-materialization-policy}
\label{sec:statefork-logical-checkpoints}
\label{sec:statefork-smart-decider}

Every call to \texttt{snapshot()} creates a logical branch point in the \sys tree.
The branch point may be represented physically or virtually.
A \emph{physical} node is backed by a checkpoint in the execution substrate.
A \emph{virtual} node stores no new backend checkpoint; instead, it records the ordered commands needed to reach that node from its nearest physical ancestor.
Restoring a virtual node first restores the nearest physical ancestor and then replays the recorded command suffix.
Thus, virtual checkpoints shift work from snapshot time and storage to restore time while preserving the same external \texttt{snapshot()}/\texttt{restore(id)} interface.

Formally, let $p(v)$ denote the nearest physical ancestor of a virtual node $v$, and let
$\mathrm{log}(p(v),v)$ be the ordered command sequence from $p(v)$ to $v$.
Then restoring $v$ is implemented as
\[
  \mathrm{restore}(v)
  =
  \mathrm{restore}(p(v));
  \prod_{c \in \mathrm{log}(p(v),v)} \mathrm{execute}(c).
\]
This definition makes materialization an internal runtime decision.
The search policy receives a checkpoint identifier from \texttt{snapshot()} and later calls \texttt{restore(id)} regardless of whether the node was represented physically or virtually.

This physical-virtual split combines checkpoints with replay logs to trade state-copying cost against recovery cost.   While this rollback/replay classic technique for containers~\citep{elnozahy2002survey,liu2018deterministic} has been applied across security~\citep{dunlap2002revirt}, debugging~\citep{srinivasan2004flashback,execRecorder2006}, and container replication~\citep{zhou2021hycor}, \sys applies this to agent exploration at a different semantic level: 
virtual nodes replay the agent's command suffix from an observation-equivalent ancestor rather than recording all nondeterministic events for exact deterministic replay.

A policy decides whether each logical node should be physical or virtual.
Physical nodes pay snapshot latency and storage overhead, but make later restores cheap.
Virtual nodes are cheap to create, but restoring them requires restoring an ancestor and replaying the suffix, whose cost grows with trajectory length.
Many terminal commands (e.g., \texttt{ls}, \texttt{cat}, \texttt{grep}, small tests) are lightweight enough to replay cheaply, avoiding the runtime and storage costs of physical checkpoints, especially for nodes that are never revisited.

\paragraph{Smart Decider} is a pragmatic and effective online checkpoint/replay policy.   It decides when to pay the cost of physical checkpoints and when to use a virtual checkpoint and replay from a materialized ancestor.  

Let $m_t$ be the resident memory size of the live session and its child processes at snapshot time $t$.
We use memory as the sole predictor because checkpoint latency for our substrate, \substrate(\S~\ref{sec:waypoint}), is largely independent of filesystem size.
From calibration measurements, \sys estimates the cost of a physical checkpoint with a simple linear model:
\[
  \widehat{C}_{\mathrm{phys}}(t)
  =
  \alpha m_t + \beta .
\]
Smart Decider supports backend-specific cost estimates that may instead depend on e.g., filesystem size, image-layer size, compression mode, VM memory, disk snapshot size, and/or export/import costs.  

Let $C_{\mathrm{replay}}(t)$ be the cumulative wall-clock command execution time since the nearest physical checkpoint.
This quantity estimates the cost of restoring the current logical node virtually, since the runtime would need to restore the nearest physical ancestor and replay that command suffix.
The Smart Decider applies the following conservative rule:
\[
  \mathrm{physical}(t)
  =
  \widehat{C}_{\mathrm{phys}}(t)
  \le
  C_{\mathrm{replay}}(t).
\]
If the estimated cost of physical checkpointing is no larger than the accumulated replay cost, \sys materializes.
Otherwise, \sys records a virtual checkpoint and keeps replay anchored at the nearest physical ancestor.

This policy has two useful properties.
First, it preserves \sys semantics: the user receives the same checkpoint identifier and uses the same \texttt{restore(id)} call independent of materialization.
Second, it avoids eagerly checkpointing every lightweight state while still creating physical checkpoints once accumulated replay cost becomes large.
The policy is intentionally simple.
It does not try to predict the full future search tree, the semantic importance of a state, or whether a branch will eventually be selected by the agent.
Predicting those quantities is a harder optimization problem that we leave to future work.
The same logical API could support richer policies, including caching frequently restored ancestors, prefetching likely restores, parallelizing branch expansion, tuning backend-specific checkpoint options, or choosing dynamically among multiple execution substrates.

\subsection{Backend Managers}
\label{sec:statefork-backends}

Backend managers translate \sys's logical operations into substrate-specific commands.
A backend manager implements the small set of physical operations needed by the session manager and materialization policy such as create and restore a physical checkpoint, execute a command, report timing and storage metadata, and clean up backend resources.
All logical tree management, command logging, and observability are handled above this backend-specific layer.
The next section presents \substrate, our execution-session-based backend optimized for terminal agents.

\section{\substrate: A Checkpoint/Restore Substrate for Terminal Agents}
\label{sec:waypoint}

While \sys provides the logical control plane for branchable exploration, \substrate is the execution substrate for terminal agents.
\substrate materializes logical branch points as local checkpoint/restore points over terminal-agent sessions.
Its checkpoint scope is the execution session itself: the filesystem, process and memory state, terminal state, and relevant local services visible to future commands.
This scope preserves the state needed for observation equivalence while excluding state unrelated to local agent exploration.

\paragraph{Design Principles}
Motivated by Section~\ref{sec:motivating_example}, \substrate follows two design principles.

First, to provide observation equivalence for terminal agents, it restores the execution session rather than an arbitrary higher or lower-level boundary.
For terminal agents, future observations depend on file contents, application memory, terminal session state, and relevant local services.
This includes application source files, installed packages, generated artifacts, working directories, environment variables, activated virtual environments, aliases, shell functions, background jobs, local servers, and the process tree that future commands may observe.
\substrate is therefore ``full-state'' with respect to the execution-session boundary, not with respect to the entire host machine.

Second, to make exploration physically efficient, it optimizes for local containment rather than portability or tenant-grade isolation.
Branches are speculative continuations of a single task on a single host, controlled by the same agent framework.
They are used to guide search, not to produce portable artifacts.
Therefore, image export, registry metadata, and container lifecycle management should not be on the common checkpoint/restore path.
They also need containment rather than a VM security boundary.
Branches are not mutually distrustful tenants, but ordinary branch effects should not leak into the host or into sibling branches.
For example, packages installed, package-manager metadata mutated, files generated, or services started in one branch must not contaminate observations in another branch unless those effects belong to a common ancestor.

\paragraph{High-level design}
A \substrate execution session, checkpoint lineage, and physical checkpoint implement their logical counterparts in \sys.
Restore returns an observation-equivalent environment for the agent's next action.  
A session contains three states: a root filesystem view, a long-lived PTY-backed shell and its descendant process tree, and checkpoint metadata that records filesystem and process-state lineage.

A tempting design would be to combine existing primitives directly by using OverlayFS for files, CRIU for processes, and a shell for commands.
This is insufficient on its own.
A filesystem snapshot alone loses shell-local state, running services, REPLs, and in-memory application state.
A raw CRIU checkpoint alone assumes that the restored process tree will see the same filesystem, mounts, PTYs, sockets, devices, and command interface.
Running every branch in a fresh container or VM captures a broader boundary, but pays for portability, lifecycle management, or machine-level state that is not needed for local speculative search.
Even the direct combination of OverlayFS and CRIU is not automatic as OverlayFS layer stacks are fixed at mount time, runtime filesystems such as \texttt{/proc} and \texttt{/sys} are not ordinary file data, and PTYs are kernel objects tied to process-session state.

\substrate coordinates these primitives at the execution-session boundary.
The manager maintains the checkpoint tree and active session root.
Each live session runs a command server attached to a PTY-backed, long-lived shell.
The shell and its descendants execute over an OverlayFS-backed root, so ordinary writes—source edits, generated files, package installations, and package-manager metadata—remain confined to the session's writable layer.
Local services started by the agent, including servers, REPLs, and databases launched from the shell, are captured with the shell's descendant process tree, including memory, descriptors, and local sockets.

Checkpointing and restoration are coordinated operations.
At checkpoint time, \substrate quiesces the command session, dumps the shell process tree with CRIU, seals the current OverlayFS upper directory as an immutable delta, records the checkpoint's parent pointer, and resumes execution over a fresh writable layer.
At restore time, \substrate terminates the current branch, rebuilds the OverlayFS lowerdir stack from checkpoint ancestry, attaches a fresh writable upperdir for the next continuation, remounts runtime filesystems, and restores the CRIU process image into the rebuilt root.
Branching is therefore a local operation over execution-session state, rather than the creation of a new container, VM, or image.

\begin{figure}[t]
  \centering
  \includegraphics[width=0.45\textwidth]{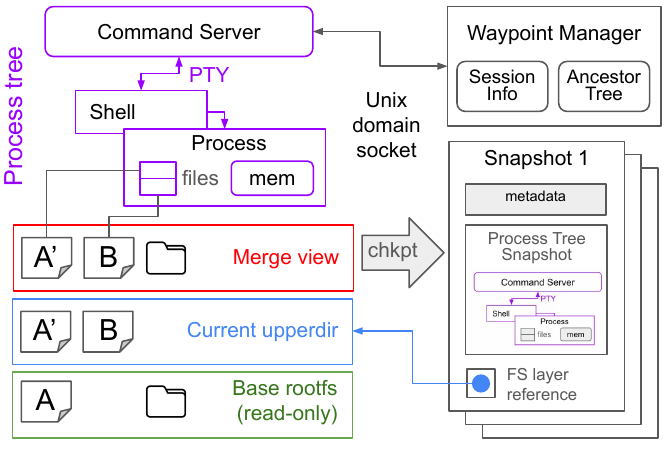}
  \caption{\textbf{\substrate architecture.}
  A live session consists of a command server, a PTY-backed long-lived shell, and process state (\textcolor{purple}{purple}) running over an OverlayFS-backed session root. The session root is exposed through the OverlayFS merge view (\textcolor{red}{red}), backed by the writable current upperdir (\textcolor{blue}{blue}) and read-only lowerdirs (\textcolor{ForestGreen}{green}). At checkpoint time, \substrate stores process state with CRIU, seals the current upperdir and records it by reference rather than copying file data, and updates checkpoint lineage metadata maintained by the manager. At restore time, \substrate rebuilds the lowerdir stack from checkpoint ancestry, attaches a fresh writable upperdir, and restores the process tree.}
  \Description{Architecture diagram of \substrate showing a manager
  connected by Unix domain sockets to live and restored session command servers.
  Each session contains a PTY-backed shell and process state over an OverlayFS
  merge view. Checkpoints store CRIU process images, metadata, and references to
  sealed filesystem layers that are reused as lowerdirs during restore.}
  \label{fig:waypoint-arch}
  \vspace{-4pt}
\end{figure}

Figure~\ref{fig:waypoint-arch} shows this architecture.
The important point is not that \substrate invents new kernel mechanisms, but that it composes existing Linux mechanisms at the boundary needed by terminal-agent exploration.

\subsection{Session-Scoped Containment}
\label{sec:waypoint-containment}

A \substrate session presents a container-like root to the agent without creating a new container object for every branch.
Commands run with their root changed to the session filesystem view and begin in \texttt{/}.
As a result, commands such as \texttt{pwd}, package managers, build scripts, and test harnesses observe paths that match the task environment rather than the host directories used to store checkpoint metadata.
This matters because terminal agents often generate absolute paths, inspect the current directory, and invoke tools whose behavior depends on the visible root.

Ordinary task writes are confined to the session.
When the agent installs packages, edits source files, writes logs, or changes configuration under the session root, those writes go into the session's writable layer.
The host stores session metadata outside the agent-visible root, while the agent sees only the task environment.
This containment model is deliberately weaker than a VM sandbox. 
\substrate does not claim to isolate malicious tenants or defend against a privileged process attempting to escape.
It targets controlled agent exploration and benchmark execution, where the primary requirement is consistent restoration of branch state.

\subsection{In-Place Filesystem Checkpointing}
\label{sec:waypoint-fs}

\substrate implements filesystem checkpointing with in-place OverlayFS layers.
Each session maintains a base root, a checkpoint parent tree, and a distinguished writable directory called \texttt{current}.
The live session runs on an OverlayFS merge view whose writable upper layer is \texttt{current}.
When a checkpoint is created, \substrate seals the current writable directory, records it under the checkpoint identifier, and allocates a fresh empty \texttt{current} directory for subsequent execution.
The sealed directory becomes an immutable delta layer.

Restoring a checkpoint reconstructs the filesystem view from checkpoint lineage.
\substrate walks the checkpoint's parent chain, orders the sealed directories as OverlayFS lower layers, appends the base root, and mounts a new merge view with a fresh writable upper layer.
Branching from a checkpoint is the same operation: (1) restore the checkpoint as the read-only ancestry, (2) attach a new writable layer, and (3) continue execution.

This  avoids the expensive path used by image-oriented container runtimes.
\substrate does not diff the merged view, copy file data, export an image layer, or synchronize a content-addressable layer database for each checkpoint.
The checkpoint stores a reference to an already-existing upper directory and a parent pointer.
Thus, installing a large package or generating many files does not make checkpoint creation proportional to the total filesystem size.
The filesystem cost is dominated by quiescing the session, unmounting and remounting the overlay, and updating \substrate metadata.

The tradeoff is that OverlayFS layer stacks are fixed at mount time.
To seal one upper directory and attach a new one, \substrate must reconstruct the overlay view.
A live process tree cannot keep executing while its root filesystem mount is torn down and replaced.
For this reason, filesystem checkpointing and process checkpointing must be coordinated, as described next.

\subsection{Stateful Shell Sessions}
\label{sec:waypoint-command-session}

A filesystem checkpoint alone is insufficient for terminal agents.
Launching a fresh shell for every command would lose shell-local state such as the current directory, exported variables, activated virtual environments, aliases, functions, and job context.
\substrate therefore makes the shell part of the checkpointed session state.

Each session contains a long-lived shell process and a small command server.
The command server listens on a Unix domain socket associated with the session.
For each command, the caller sends a complete command payload; the server forwards the command to the persistent shell and returns the captured output.
This gives agents a simple command-step interface while preserving the shell state accumulated across previous commands.

The shell is connected through a pseudo-terminal (PTY) rather than ordinary pipes.
Many command-line tools change behavior depending on whether they are attached to a terminal.
A PTY gives the shell terminal-compatible behavior while still exposing a programmatic request/response interface to the agent.
\substrate does not attempt to expose a full interactive terminal.
It targets the command-step model used by terminal-agent benchmarks: the agent submits a complete command that runs to completion or times out; the observation outputs to the agent.
Full-screen terminal applications, arbitrary signal delivery, live editing, and complex interactive job control are outside this interface.

The command protocol preserves command boundaries.
Each request carries one command.
The server writes the command to the PTY, appends a unique completion marker, drains PTY output, and returns output only after the marker appears.
Before returning, \substrate removes shell artifacts such as echoed commands, prompts, markers, and terminal control sequences.
Command execution is serialized within a session, so two requests cannot interleave on the same PTY.

\subsection{Process and Memory Checkpointing}
\label{sec:waypoint-process}

\substrate uses CRIU to capture the process state of the session.
When a session starts, \substrate records the PID of the long-lived shell.
To create a process checkpoint, \substrate invokes CRIU on that shell PID.
Because agent commands execute as children of the shell, dumping the shell process tree captures the shell, foreground commands, background jobs, local servers, REPLs, and other descendant processes that are part of the session.

A checkpoint therefore contains two coordinated components.
The filesystem component is the sealed OverlayFS upper directory and its checkpoint lineage.
The process component is the CRIU image directory for the shell process tree.
During checkpoint creation, \substrate first quiesces the session and asks CRIU to dump the shell process tree.
The CRIU images are stored under the checkpoint's directory.
While the process tree is stopped, \substrate seals the current filesystem layer, records the checkpoint metadata, creates a fresh writable layer, reconstructs the overlay view, and restores the shell process tree into that view.

Restoring an earlier checkpoint follows a similar process.
\substrate terminates the currently running shell process tree, reconstructs the overlay view corresponding to the target checkpoint, and invokes CRIU restore using the process images stored with that checkpoint.
The restored shell resumes as the session shell, with its process state, shell-local variables, working directory, and child process state restored as part of the CRIU image.
In normal operation, the session shell is either the original shell created at session initialization or a CRIU-restored instance of that shell; \substrate does not replace it with a fresh shell after every restore.

This design makes CRIU useful at the right boundary.
Raw CRIU does not know how to reconstruct the session filesystem tree, and filesystem checkpoints do not preserve memory.
\substrate combines the two: durable file changes are represented as sealed OverlayFS layers, while shell context and application memory are represented as process state.

\subsection{Composing Primitives}
\label{sec:waypoint-composition}

The main implementation challenge is that the Linux mechanisms used by \substrate do not naturally compose into one checkpoint abstraction.
OverlayFS constructs a merged directory tree from upper and lower directories.
CRIU restores a process/resource graph.
Runtime filesystems such as \texttt{/proc} and \texttt{/sys} are kernel-provided views rather than ordinary files.
PTYs are stateful kernel objects tied to open descriptors, terminal peers, and process-session state.
Each is useful individually, but their boundaries do not align.

\paragraph{Runtime mounts.}
Runtime filesystems cannot simply be mounted once inside a lower layer and expected to appear correctly through every future overlay merge view.
\substrate instead treats them as part of the active session root.
After reconstructing the overlay, it mounts the required runtime filesystems under the active root; before tearing down the overlay, it detaches them.
This keeps ordinary file contents in OverlayFS layers while keeping kernel-provided views tied to the currently active session.

\paragraph{PTYs and \texttt{devpts}.}
PTYs introduce a separate composition problem.
In the Unix PTY model, slave devices are exposed through \texttt{/dev/pts}, and different \texttt{devpts} mounts can represent different PTY namespaces.
Making \texttt{devpts} part of every reconstructed overlay root would make terminal identity depend on the changing mount topology that CRIU must restore.
\substrate therefore allocates PTYs through the host \texttt{devpts} instance while running the child shell inside the session root.
This preserves terminal-compatible shell behavior without introducing a nested \texttt{devpts} mount into the overlay layer stack.

\paragraph{Overlay reconstruction and process state.}
The filesystem and process mechanisms also conflict around liveness.
\substrate obtains filesystem-size-independent checkpoints by sealing the current upper directory and remounting a new overlay stack.
However, a process cannot continue executing normally while its root filesystem view is being unmounted and replaced.
Consequently, \substrate cannot simply use CRIU's leave-running style of process checkpointing across filesystem snapshots.
It prioritizes filesystem consistency and stops the session process tree, reconstructs the filesystem view, remounts runtime filesystems, and restores the process tree into the new view.

\section{Evaluation}
\label{sec:eval}
We evaluate \sys and \substrate using both microbenchmarks and macrobenchmarks. 
The microbenchmarks isolate the cost of snapshot and restore operations under controlled memory and file system state sizes, allowing us to examine how efficient different substrates are as the preserved state grows.
The macrobenchmarks instead evaluate end-to-end agent exploration workloads, where restoration cost and restoration fidelity together determine whether agents can efficiently branch, resume, and continue exploration from intermediate execution states. 

\subsection{Microbenchmark}
\label{sec:microbenchmark}
\task{Primary: Alex, Danielle.}




\begin{figure*}[t]
    \centering
    \includegraphics[width=\textwidth]{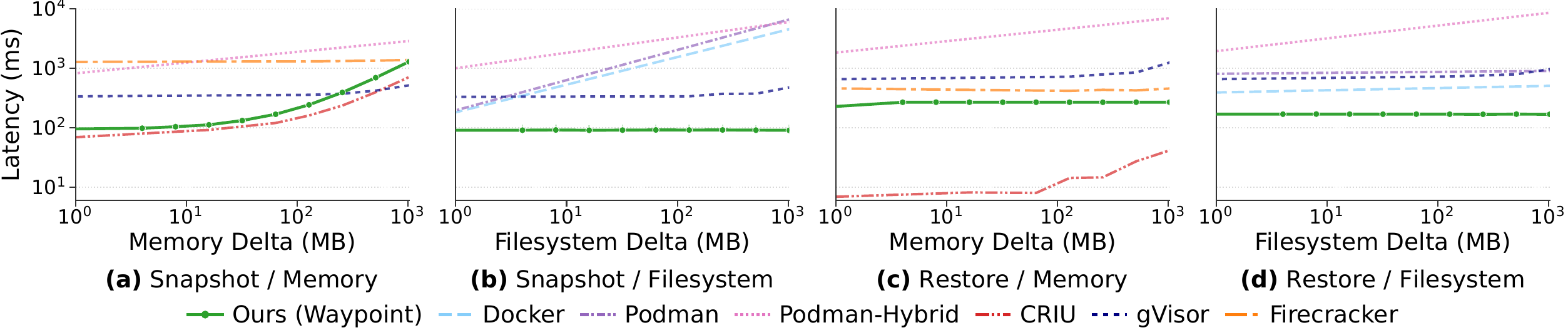}
    \caption{Snapshot and restore latency under increasing memory and file system sizes.}
    \label{fig:microbenchmark-snapshot-restore}
\end{figure*}

We compare the snapshot and restore latency of \sys, CRIU, Podman+CRIU hybrid, gVisor, and Firecracker under different memory and file system sizes.
We used a test workload with configurable memory and disk usage, running and checkpointing this workload through the \sys interface for a consistent comparison between substrates.
To measure snapshot latency, for each setting of the memory/disk size, we initialized a new environment using \sys, ran the test workload, and then took multiple snapshots back-to-back, recording the median elapsed time.
To measure restore latency, for each setting of the memory/disk size, we initialized a new environment using \sys, took an initial snapshot, ran the test workload, and then took another snapshot.
Then, we repeatedly restored to the base snapshot and then back to the workload snapshot, recording the median elapsed time for the workload restores. 

We run the microbenchmarks on a CloudLab \texttt{c6620} server with a 28-core Intel Xeon Gold 5512U CPU at 2.1~GHz, 128~GB ECC memory, and an 800~GB Dell DC NVMe ISE 7450 MU U.2 SSD.
The machine runs Ubuntu 24.04.3 LTS with Linux 6.8.0-71-generic, and the host filesystem is ext4.
The software versions are Docker 28.2.2, Podman 4.9.3, \texttt{runc} 1.3.3, \texttt{runsc} 1.1.0-rc.1, and CRIU 4.2.0.

We evaluated each substrate on the state supported by its checkpointing mechanism.
For memory state, we compared Firecracker, CRIU, the Podman+CRIU hybrid, gVisor, and \substrate; for file system state, we compared Docker, Podman, gVisor, and \substrate. Figures \ref{fig:microbenchmark-snapshot-restore}(a) and \ref{fig:microbenchmark-snapshot-restore}(c) compare snapshot and restore latency, respectively, as memory usage increases, while Figures \ref{fig:microbenchmark-snapshot-restore}(b) and \ref{fig:microbenchmark-snapshot-restore}(d) compare latency across file system sizes.

The memory and disk state sizes ranged from 0 MB to 1024 MB each.
Firecracker allows the user to configure the size of the microVM memory, so if the user knows in advance the maximum workload size, choosing a smaller memory will enable faster snapshots.
In these experiments, we configured the microVM memory size to 2 GB, accommodating the largest workload tested.

In the memory microbenchmark, \substrate's snapshot latency closely follows bare CRIU because memory capture is dominated by CRIU's process dump, while \substrate adds only lightweight file system-layer bookkeeping.
Across the resident-memory range relevant to our terminal-agent workloads, \substrate is lower latency than the other substrates that capture both memory and file system state.
Only at very large synthetic resident-memory sizes does \substrate approach or exceed some baselines with mostly fixed snapshot costs.
In particular, gVisor, by implementing its own userspace kernel, does not need to use CRIU and can be more efficient for large memory sizes.
We do not observe this regime in the relevant benchmarks.
For restore, \substrate is slower than bare CRIU because it restores more than process memory.
It also reconstructs the OverlayFS view, attaches a fresh writable layer, remounts runtime filesystems, and resumes the process tree inside the restored execution session.
This extra work is necessary for observation equivalence because CRIU alone does not restore the file system/session boundary seen by future commands.
Despite this broader state boundary, \substrate's restore latency remains consistently lower than the evaluated container and VM baselines across the memory range.

In the filesystem microbenchmark, \substrate decouples snapshot and restore latency from filesystem size.
Other evaluated substrates scale with the amount of filesystem state because they copy, diff, package, or reconstruct file data on the checkpoint/restore path.
\substrate instead records filesystem checkpoints by sealing the current OverlayFS upper directory and later restores by rebuilding the OverlayFS mount stack, so the dominant work is remounting metadata rather than moving file contents.
\substrate scales well to large filesystems, we find that for {\it any} file system size evaluated, \substrate has the lowest latency, with snapshot latency $3.6\times$ lower and restore latency $2.1\times$ lower than the best competing substrate.
This shows that \substrate's advantage comes both from filesystem-size independence and from avoiding container- or VM-level machinery on the common checkpoint/restore path.

\subsection{Macrobenchmark}
\label{sec:macrobenchmark}
\task{Primary: Ruizhe, Tianle. Support: Alex.}
\noindent\textbf{Workload.} We evaluate how different execution substrates satisfy the two requirements identified in
\S~\ref{sec:physical-requirements}: observation-equivalent restoration of intermediate nodes and physical efficiency.
For our end-to-end test, we use Terminal-Bench~\cite{tbench_2025} as the representative workload for terminal-use agents.
To systematically examine how \sys-enabled exploration contributes to task completion, we instantiate two classes of agent strategies: pass@$20$, and Monte Carlo Tree Search (MCTS).
To make the methods comparable, we match the effective exploration budget across setups.
All MCTS variants use the same \sys API and search policy without the Smart Decider optimization; they differ only in how a logical checkpoint is materialized.

\noindent\textbf{Baseline Execution Substrates.}\emph{Prefix Replay} implements checkpoints virtually; to restore a node, it resets the task environment and re-executes the command prefix leading to that node.
\emph{Docker commit} snapshots the container file system layer.
\emph{Podman-Hybrid} combines CRIU checkpointing with filesystem state, capturing a broader boundary than Docker commit.
The macrobenchmark requires a backend to expose a programmatic create/restore/delete interface, support repeated branch restoration for Terminal-Bench-compatible environments, and preserve the agent-visible task interface under identical search control flow.
Systems such as gVisor and Firecracker are useful points in the broader design space, but they are not comparable drop-in macrobenchmark baselines for this experiment.
gVisor primarily changes the sandbox boundary and introduces runtime and networking constraints around checkpoint/restore~\citep{gvisor,moby-issue,gvisor-issue}, while  Firecracker snapshots~\citep{firecracker,firecracker-snap} operate at a VM boundary and require separate coordination of guest memory, disk state, and task setup while leading to higher latencies as shown in the previous section.

\noindent\textbf{Setup.}We run the macrobenchmark using gpt4.1mini with the substrates hosted on AWS EC2 \texttt{i4i.xlarge} instances, which provide 4~vCPUs, 32~GB of memory, and one 937~GB local AWS Nitro SSD~\cite{aws-i4i}.
Each instance runs Ubuntu 24.04.3 LTS with Linux 6.17.0-1009-aws.

\begin{figure*}[t]
    \centering
    \includegraphics[width=\textwidth]{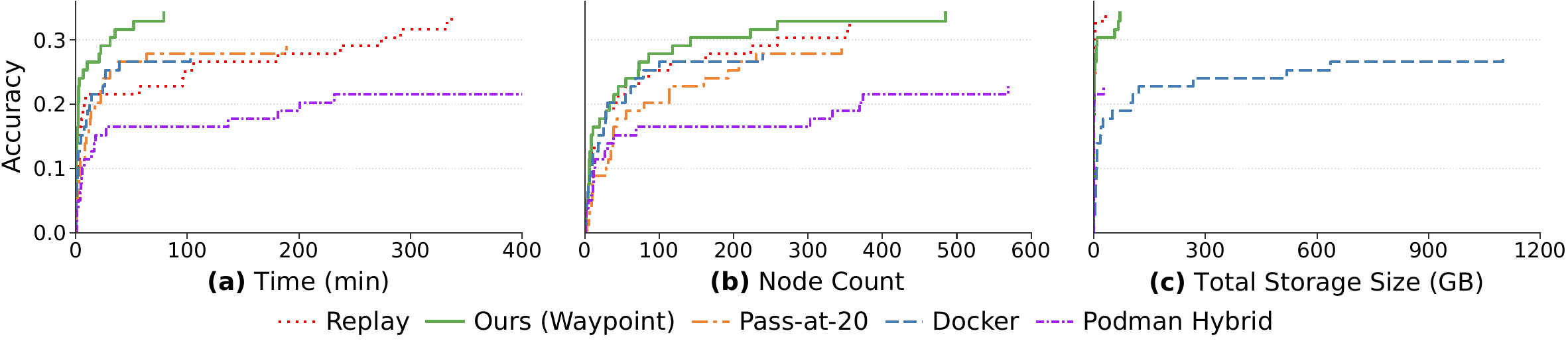}
    \caption{Terminal-Bench task completion rate as a function of wall-clock exploration time, nodes explored, and storage used.}
    \label{fig:terminal-bench-accuracy-combined}
\end{figure*}

\paragraph{Experimental Results.}
Figure~\ref{fig:terminal-bench-accuracy-combined}(a) shows task completion rate as a function of wall-clock exploration time.
\sys with \substrate reaches the best accuracy--time trade-off because it restores the execution-session state needed for observation equivalence without replaying long prefixes or paying deployment-oriented container overheads.
Prefix Replay reaches the same accuracy because it reconstructs the necessary state by re-executing commands, but it is much slower as restore cost accumulates with trajectory length.
Docker commit improves over time but wastes search effort on nodes whose restored state is incomplete.
Podman-Hybrid captures more state than Docker commit, but its higher overhead and failures around resources such as sockets and local services limit progress.

Figure~\ref{fig:terminal-bench-accuracy-combined}(b) compares sample efficiency in terms of explored nodes.
\substrate achieves higher accuracy with fewer nodes than Docker commit and Podman-Hybrid, indicating that restored nodes more consistently preserve the execution context needed for useful continuation.
Prefix Replay is similar in node efficiency because it can reconstruct the correct state, but, as shown in Figure~\ref{fig:terminal-bench-accuracy-combined}(a), it pays substantially higher wall-clock time to do so.
Docker commit eventually improves as more nodes are explored, but its lower accuracy at the same node count suggests that many samples are spent from states that are not observation-equivalent to the intended search node.
Podman-Hybrid saturates earlier because some checkpoints fail or become expensive when tasks involve resources such as sockets and local services.


Figure~\ref{fig:terminal-bench-accuracy-combined}(c) reports the cumulative checkpoint sizes.
Docker stores container-sized artifacts and incurs considerable costs, while Podman-hybrid compresses its artifacts which saves storage but incurs considerable checkpoint latency---even so, they don't capture the state needed for observation equivalence, so have low accuracy.
In contrast, \substrate checkpoints execution-session state that incurs modest storage but much higher accuracy.
Replay only requires storage for one docker image, but is penalized in restore time.

Overall, these results show that effective terminal-agent exploration requires both sufficient execution-session state and low-cost restoration.
\sys provides the branchable exploration API, while \substrate provides the backend that restores the right state efficiently.

\begin{figure}
    \centering
    \includegraphics[width=1\linewidth]{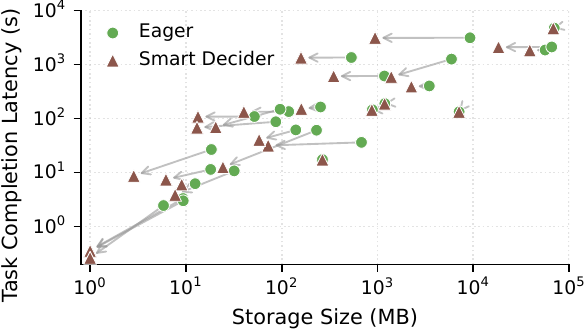}
    \caption{Smart Decider (brown triangles) reduces storage size and often task latency as compared to Eager (green circles).  Each point is a task, and arrows point from Eager to Smart Decider.  
    }
    \label{fig:smart-decider-storage-time}
    \vspace{-4pt}
\end{figure}

Figure~\ref{fig:smart-decider-storage-time} uses \sys with \substrate to compare Eager materialization with Smart Decider, which uses virtual branch points and reconstructs via replay from the closest materialized ancestor.   The task actions are identical, and differences reflect how \sys physically represents the same logical exploration tree.  

The Smart Decider reduces physical exploration cost substantially.
Across tasks, it lowers completion time by 23\% on average while reducing checkpoint storage by 58\% relative to eager \sys+\substrate materialization.
These aggregate improvements come from avoiding checkpoints that are cheap to represent logically but expensive to materialize physically.
The largest gains occur on tasks where eager materialization creates many intermediate checkpoints that are never restored or for actions that are very cheap to execute.
In those cases, selective materialization avoids unnecessary snapshot cost and storage growth while preserving the same external \texttt{snapshot}/\texttt{restore} semantics.

\section{Discussion and Limitations}
\label{sec:discussion}

\textbf{Containment, not sandboxing.}
\substrate is designed for branch isolation within one agent task, not for mutually distrustful tenants.
Its session-scoped containment prevents ordinary task writes from leaking into the host-visible environment or into other branches, but it is deliberately weaker than a VM security boundary.
A malicious process running with sufficient privileges inside the session may still exploit kernel interfaces or misconfigured mounts.
This is appropriate for our target setting---controlled agent exploration and benchmark execution---but production deployments with arbitrary untrusted code should combine \substrate with stronger sandboxing mechanisms.

\textbf{External and distributed state.}
\sys restores local state 
but tasks that depend on external services, remote APIs, or network peers outside the session root may require additional coordination. 
Some workloads can address this by running auxiliary services inside the same session or by using task-specific reset hooks, but \sys does not attempt to provide a general distributed snapshot mechanism.
This limitation is shared by many practical checkpointing systems: preserving observation equivalence depends on whether all task-relevant state is inside the checkpoint boundary.


\paragraph{GUI}
\substrate currently targets terminal agents through a PTY-backed shell and text  observations.
Adding GUI support would mainly require widening the execution-session boundary to include a headless display server, window manager, GUI application processes, screenshot service, as well as mouse/keyboard input proxy.
These components could run inside the same OverlayFS-backed session and be captured as part of the checkpointed process tree, leaving the \sys branch, restore, and cleanup API unchanged.
Thus, headless software-rendered GUI support is a natural extension of \substrate, while full desktop or hardware-accelerated GUI exploration would require broader device and display-server checkpointing than the current terminal-focused implementation.

\section{Conclusion}
\label{sec:conclusion}

Agent exploration is most effective when agents can reuse intermediate nodes in a search tree, but doing so for terminal-use agents requires restoring the execution-session state that determines future observations.
This paper introduced \sys, a logical control plane that separates agent search from physical state materialization, and \substrate, a checkpoint/restore substrate that efficiently captures filesystem, process, memory, and terminal-session state.
Our results show that branch-based exploration improves sample efficiency, while execution-session checkpointing and selective materialization reduce the physical cost of realizing that exploration.
Together, \sys and \substrate provide efficient infrastructure for exploratory AI agents.

\bibliographystyle{plain}
\bibliography{main}

\newpage
\newpage
\appendix
\section{Experimental Setup Details}

\subsection{Search Algorithms}
\paragraph{Pass@{\boldmath$20$}.} 
The pass@$20$ baseline relaxes the one-shot constraint by allowing the agent to interact iteratively with the system. Specifically, in each trial the agent may issue a sequence of commands, observe the resulting state and outputs, and adapt subsequent actions accordingly. The process is repeated for at most $20$ independent trials, with success recorded if any single trial reaches the target state. This is a common baseline in many coding/system agent benchmarks \citep{xie2024osworldbenchmarkingmultimodalagents, jimenez2024swebench, tbench_2025}, where agents are allotted a small budget of retries.
For pass@$20$, the agent is allowed $n=30$ exploratory rounds per trial with a total of two iterations.

\paragraph{Monte Carlo Tree Search (MCTS).} 
To capture structured exploration beyond simple retries, we model multi-step interaction using Reflective Monte Carlo Tree Search. At each \emph{environment step} of task execution, the agent initiates a tree search simulation with branching factor $w$ and depth $d$, simulating alternative continuations of the current trajectory. Each rollout explores a possible command sequence, and search nodes are evaluated using the intermediate system observations as feedback signals. After the search completes, an LLM-based decision module selects the best move to commit at the current environment step. Execution then proceeds to the next environment step, where the process repeats. The process terminates when either the agent declares the task complete or the overall budget of $k$ main steps has been exhausted.
For the MCTS-based methods, we fix the branching factor to $w=3$ and search depth to $d=2$ at each environment step.

\paragraph{Variance and replication.}
Each configuration is evaluated in a single experimental run, that run already aggregates substantial sampling: every condition involves hundreds of stochastic LLM calls (20 independent rollouts per task under pass@20, and a matched visited-node budget under MCTS), and reported accuracy averages over all 77 Terminal-Bench tasks. 

\end{document}